\documentclass[a4paper,11pt]{article}
\pdfoutput=1
\usepackage{a4wide}

\usepackage{amsmath}
\usepackage{amsfonts}
\usepackage{amssymb}
\usepackage{amsthm}
\usepackage{latexsym}
\usepackage{bm}
\usepackage{graphicx}
\usepackage[colorlinks]{hyperref}

\newcommand{\ie}[0]				{i.e.}

\newcommand{\cf}[0]				{\textit{cf.}}

\newcommand{\shalf}[0]			{\tfrac 1 2}

\newcommand{\bx}				{{\boldsymbol x}}
\newcommand{\by}				{{\boldsymbol y}}
\newcommand{\bz}				{{\boldsymbol z}}

\newcommand{\bX}				{{\boldsymbol X}}
\newcommand{\bY}				{{\boldsymbol Y}}
\newcommand{\bZ}				{{\boldsymbol Z}}
\newcommand{\bU}				{{\boldsymbol U}}
\newcommand{\bxi}				{{\boldsymbol\xi}}
\newcommand{\beeta}				{{\boldsymbol\eta}}
\newcommand{\btheta}			{{\boldsymbol\theta}}
\newcommand{\tq}				{\tilde{q}}
\newcommand{\hp}				{{\hat p}}

\newcommand{\hbtheta}			{{\hat\btheta}}

\newcommand{\veps}				{\varepsilon}
\newcommand{\vphi}				{\varphi}
\newcommand{\sbtheta}			{{\btheta^*}}
\newcommand{\hSigma}			{{\widehat\Sigma}}

\newcommand{\aslim}				{\,\xrightarrow{\,a.s.}\,}

\newcommand{\probop}			{{\mathbf P}}
\newcommand{\expectop}			{{\mathbf E}}
\newcommand{\entrop}			{{\mathbf H}}

\DeclareMathOperator{\likel}	{\mathcal{L}}

\DeclareMathOperator{\allem}	{\hat\ell}

\DeclareMathOperator{\TE}		{\mathcal{T}}
\DeclareMathOperator{\TEE}		{\widehat{\TE}}

\newcommand{\te}[2]				{\TE_{{#1}\to{#2}}}
\newcommand{\cte}[3]			{\TE_{{#1}\to{#2}|{#3}}}

\newcommand{\tee}[3]			{\TEE_{{#1}\to{#2}}\bracr{#3}}

\newcommand{\bracr}[1]			{\left({#1}\right)}

\newcommand{\lcond}[2]			{\left.{#1}\,\right|{#2}}

\newcommand{\prob}[1]			{\probop\!\bracr{#1}}

\newcommand{\expect}[1]			{\expectop\!\bracr{#1}}

\newcommand{\entro}[1]			{\entrop\!\bracr{#1}}
\newcommand{\centro}[2]			{\entro{\lcond{#1}{#2}}}

\DeclareMathOperator{\TEF}		{{\TEE\!\raisebox{4pt}{$\scriptstyle p$}\,}}
\newcommand{\tefna}[2]			{\TEF_{{#1}\to{#2}}}

\newcommand{\figref}[1]			{FIG.~\ref{#1}}

\newcommand{\resub}[1]			{{#1}}

\theoremstyle{plain}
\newtheorem{temlprop}{Proposition}
\newtheorem{temltheo}{Theorem}

\begin{document}

\title{Transfer Entropy as a Log-likelihood Ratio}

\author{Lionel Barnett\footnote{\href{mailto:l.c.barnett@sussex.ac.uk}{l.c.barnett@sussex.ac.uk}} \\
	\small{Sackler Centre for Consciousness Science} \\
	\small{School of Informatics} \\
	\small{University of Sussex} \\
	\small{Brighton BN1 9QJ, UK} \\
	\\
	Terry Bossomaier\footnote{\href{mailto:tbossomaier@csu.edu.au}{tbossomaier@csu.edu.au}} \\
	\small{Centre for Research in Complex Systems} \\
	\small{Charles Sturt University} \\
	\small{Panorama Ave, Bathurst NSW 2795, Australia}
}

\date{\today}

\maketitle

\begin{abstract}
Transfer entropy, an information-theoretic measure of time-directed information transfer between joint processes, has steadily gained popularity in the analysis of complex stochastic dynamics in diverse fields, including the neurosciences, ecology, climatology and econometrics. We show that for a broad class of predictive models, the log-likelihood ratio test statistic for the null hypothesis of zero transfer entropy is a consistent estimator for the transfer entropy itself. For finite Markov chains, furthermore, no explicit model is required. In the general case, an asymptotic $\chi^2$ distribution is established for the transfer entropy estimator. The result generalises the equivalence in the Gaussian case of transfer entropy and Granger causality, a statistical notion of causal influence based on prediction via vector autoregression, and establishes a fundamental connection between directed information transfer and causality in the Wiener-Granger sense.
\end{abstract}

\maketitle


Transfer entropy (TE) was formulated by Schreiber \cite{Schreiber00} as a non-parametric measure of directed (time-asymmetric) information transfer between joint processes. It has since rapidly gained popularity, particularly in neuroscience \cite{Hinrichs:2006,Gourevitch:2007,Vakorin:2010,Vicente:2011,Lizier:2011}, as a tool for data-driven detection of functional coupling between joint processes. In \cite{Barnett:2009} it was shown that for Gaussian vector autoregressive (VAR) processes, transfer entropy is equivalent to Granger causality \cite{Wiener:1956,Granger63,Geweke82}. Noting that the Granger causality statistic may be formalised as a log-likelihood ratio, this Letter extends the result in \cite{Barnett:2009} (see also \cite{Amblard:2011,Seghouane:2012}) to a very general class of continuous or discrete Markov models in a maximum likelihood framework. In the case of a finite state space, moreover,  no explicit model is required since the result applies to the conventional plug-in estimator for TE.

The result is of particular significance since estimation of TE in sample---particularly for continuous systems---is notoriously awkward \cite{Schreiber00,KaiserSchreiber02,RungeEtal:2012}; furthermore, little has been known previously about its sampling properties. The likelihood formulation presented here provides a convenient route to estimation and statistical inference of TE in a broad parametric context, and on a conceptual level pinpoints the relationship between information-theoretic directed information transfer in the Schreiber sense and predictive, model-based causality in the Wiener-Granger sense.


Firstly we introduce some notation \footnote{In what follows upper case symbols denote random variables and bold typeface indicates vector quantities.}: for a time-indexed sequence $\bx = (x_t)$, $t = 1,2,\ldots$ we write $\bx^t   = (x_1,x_2,\ldots,x_t)$ for the subsequence up to time $t$, and $\bx^k_t = (x_{t-k},\ldots,x_{t-1})$ for the $k$-lag history of the sequence up to time $t-1$. Suppose now that $(\bX,\bY) = (X_t,Y_t)$ are jointly stationary stochastic processes taking values in the state space $S_X \times S_Y$. (Later we shall also require an \emph{ergodicity} condition on the joint process.) We then have the following expressions for the entropy of $X_t$ conditional on the joint ($k$-lag) history of $\bX$ and $\bY$, and on the history of $\bX$ only:
\begin{equation}
\begin{split} \label{eq:entrodef}
	\centro{X_t}{\bX^k_t,\bY^k_t} &= -\expect{\log p(X_t | \bX^k_t,\bY^k_t)} \\
	\centro{X_t}{\bX^k_t} &= -\expect{\log p(X_t | \bX^k_t)}
\end{split}
\end{equation}
(assuming that the expectations are $< \infty$) where $p(x_t|\bx^k_t,\by^k_t)$, $p(x_t|\bx^k_t)$ are respectively the marginal probability density functions (pdfs) of $X_t$ conditioned on the history of $\bX$ and $\bY$, and on the history of $\bX$ only. By stationarity, these densities do not depend on $t$. The ($k$-lag) transfer entropy from $Y \to X$ is then defined by:
\begin{equation}
	\te Y X \equiv \centro{X_t}{\bX^k_t} - \centro{X_t}{\bX^k_t,\bY^k_t} \,, \label{eq:trent}
\end{equation}
which may be read as ``the degree to which the history of $\bY$ disambiguates current $\bX$ beyond the degree to which $\bX$ is already disambiguated by its own history".

Suppose now given a parametrised Markov (predictive) model for $X_t$ in terms of the history of $\bX$ and $\bY$:
\begin{equation}
	p(x_t | \bx^{t-1}, \by^{t-1}; \btheta) = f(x_t|\bx^k_t,\by^k_t;\btheta) \,, \label{eq:parmod}
\end{equation}
where $\btheta = (\theta_1,\ldots,\theta_m)$ is a vector of parameters in a (Euclidean) parameter space $\Theta$, $f(\cdot|\cdot,\cdot;\btheta)$ is a conditional probability density function from $S_X \times S^k_X \times S^k_Y \to [0,1]$ and $p(x_t | \bx^{t-1}, \by^{t-1}; \btheta)$ is the marginal pdf of $X_t$ conditioned on the \emph{entire} history of $\bX$ and $\bY$ under the model assumption with parameter vector $\btheta$. Eq.~\eqref{eq:parmod} describes a ``partial'' model, insofar as only the marginal conditional distribution of $\bX$ is specified. We assume that the model is \emph{identifiable} and \emph{well-specified}: that is, $\btheta_1 \ne \btheta_2 \implies f(\cdot|\cdot,\cdot;\btheta_1) \ne f(\cdot|\cdot,\cdot;\btheta_2)$, and there is a unique \emph{true} parameter vector $\sbtheta$ satisfying
\begin{equation}
	p(x_t|\bx^{t-1},\by^{t-1}) = f(x_t|\bx^k_t,\by^k_t;\sbtheta) = p(x_t|\bx^k_t,\by^k_t) \,. \label{eq:parmodt}
\end{equation}
We do not demand that the \emph{joint} process $(\bX,\bY)$ satisfy a Markov property analogous to \eqref{eq:parmodt}.

To apply likelihood methods we require a ``full'' rather than a partial, model \footnote{An alternative approach would be to regard the $y_t$ as \emph{nuisance parameters}; however, we believe this complicates the analysis.}. However, we are not really interested in modelling $\bY$, but just in how its history ``drives'' $\bX$. As a mathematical device, then, we extend the partial model (somewhat arbitrarily) to a full model, for which the  marginal distribution of $X_t$ conditional on joint history agrees with \eqref{eq:parmod}. Let $q(y)$ be the marginal (unconditional) pdf of $Y_t$. We define the extended model (again parametrised by $\btheta \in \Theta$) by \footnote{In fact, as regards the subsequent argument, the marginal $Y_t$ pdf $q(y)$ in \eqref{eq:fullmod} could be replaced by an arbitrary pdf $\tq(y)$ with $\tq(y) \ne 0$ almost everywhere.}
\begin{equation}
	p(x_t,y_t | \bx^{t-1}, \by^{t-1}; \btheta) = f(x_t|\bx^k_t,\by^k_t;\btheta) q(y_t) \,, \label{eq:fullmod}
\end{equation}
where $p(x_t,y_t | \bx^{t-1}, \by^{t-1}; \btheta)$ is the conditional \emph{joint} pdf of $X_t,Y_t$ on their joint history under the model assumption with parameter vector $\btheta$. Eq.~\eqref{eq:fullmod} may be interpreted as ``the $Y_t$ are independent of $X_t$ conditioned on joint history''. Note that in fact this condition will in general not actually hold for the joint process $(\bX,\bY)$; \ie\ in general there will be no $\btheta \in \Theta$ such that $p(x_t,y_t | \bx^{t-1}, \by^{t-1}) = f(x_t|\bx^k_t,\by^k_t;\btheta) q(y_t)$ [\cf\ \eqref{eq:parmodt}]. In other words, the model \eqref{eq:fullmod} will generally be \emph{misspecified}. As we shall see, however, this is not problematic.

Misspecified or not, there is nothing preventing calculation of \emph{likelihoods} for the extended model \eqref{eq:fullmod}. Suppose given a realisation $(\bx^n,\by^n)$ of length $n$ sampled from the joint process $(\bX,\bY)$. The \emph{likelihood} of $\btheta$ given this realisation is defined as
\begin{equation}
	\likel(\btheta|\bx^n,\by^n) \equiv p(\bx^n, \by^n; \btheta) \,,
\end{equation}
\ie\  the pdf of the joint history $(\bX^n,\bY^n)$ under the model assumption with parameter vector $\btheta$, and the \emph{average log-likelihood estimator} is defined to be
\begin{equation}
	\allem(\btheta|\bx^n,\by^n) \equiv \frac 1{n-k} \log \likel(\btheta|\bx^n,\by^n) \label{eq:allem}
\end{equation}
(the factor of $n-k$ is due to the effective loss of $k$ samples due to lags). We have by Bayes' theorem
\begin{align*}
	\likel(\btheta|\bx^n,\by^n)
	&= p(x_n,y_n | \bx^{n-1}, \by^{n-1}; \btheta) p(\bx^{n-1}, \by^{n-1}; \btheta) \\
	&= f(x_n|\bx^k_n,\by^k_n;\btheta) q(y_n) \likel(\btheta|\bx^{n-1},\by^{n-1})
\end{align*}
from \eqref{eq:parmod}, leading to
\begin{equation}
	\likel(\btheta|\bx^n,\by^n) = p(\bx^k,\by^k) \prod_{t = k+1}^n [f(x_t|\bx^k_t,\by^k_t;\btheta) q(y_t)] \label{eq:likel}
\end{equation}
where $p(\bx^k,\by^k)$, the initial joint distribution, is assumed given independently of $\btheta$. Noting that the $q(y_t)$ do not reference the parameter vector $\btheta$, we see that \emph{up to an additive factor not depending on $\btheta$}---which does not affect calculation of either maximum likelihood (ML) estimates or likelihood ratios---and assuming $q(y) \ne 0$ almost everywhere (which in fact follows from the ergodic assumption introduced below), the average log-likelihood estimator is just
\begin{equation}
	\allem(\btheta|\bx^n,\by^n) = \frac 1{n-k} \sum_{t = k+1}^n \log f(x_t|\bx^k_t,\by^k_t;\btheta) \,. \label{eq:alle1}
\end{equation}
In practice this expression may be used to estimate an appropriate model order (\ie\ number of lags $k$) via standard likelihood-based techniques such as the Akaike or Bayesian Information Criteria \cite{McQuarrie98}.

We now make the following \emph{ergodic assumption}:
\begin{equation} \label{eq:ergodic}
	\text{The process } \bU \textrm{ is ergodic, where } U_t \equiv (X_{t-k},\ldots,X_t,Y_{t-k},\ldots,Y_{t-1}) \,.
\end{equation}
The Birkhoff-Khinchin ergodic theorem \cite{Billingsley65} then applies, so that \footnote{$\aslim$ denotes almost sure convergence.}
\begin{equation}
	\allem(\btheta|\bX^n,\bY^n) \aslim \expect{\log f(X_t | \bX^k_t,\bY^k_t;\btheta)} \label{eq:alle}
\end{equation}
as $n \to \infty$, again assuming that the expectation exists (at least for almost all $\btheta$). In particular, for the \emph{true} parameter $\sbtheta$ it follows from \eqref{eq:parmodt} and the definition \eqref{eq:entrodef} that
\begin{temlprop} \label{prop:allentro}
\begin{equation}
	\allem(\sbtheta|\bX^n,\bY^n) \aslim -\centro{X_t}{\bX^k_t,\bY^k_t} \label{eq:alogl}
\end{equation}
as $n \to \infty$. \qed
\end{temlprop}
\noindent \resub{Proposition~\ref{prop:allentro} encapsulates the relationship between average log-likelihood and conditional entropy for a Markov model, which is key to our main result.} Now by Gibbs' Inequality, we have $\expect{\log f(X_t | \bX^k_t,\bY^k_t;\btheta)} \le \expect{\log p(X_t | \bX^k_t,\bY^k_t)}$ for all $\btheta$, with equality iff $f(\cdot|\cdot,\cdot;\btheta) = p(\cdot|\cdot,\cdot)$ almost everywhere. By  uniqueness of $\sbtheta$, $\btheta = \sbtheta$ maximises $\allem(\btheta|\bx^n,\by^n)$ almost surely in the limit $n \to \infty$ and we have
\begin{temlprop} \label{prop:mle}
\begin{equation}
	\hbtheta(\bX^n,\bY^n) \aslim \sbtheta \label{eq:mle}
\end{equation}
as $n \to \infty$, where $\hbtheta(\bX^n,\bY^n)$ is the ML estimator for the extended model \eqref{eq:fullmod}. \qed
\end{temlprop}
\noindent Thus, notwithstanding that the extended model may be misspecified, its ML estimator $\hbtheta(\bX^n,\bY^n)$ is nonetheless a consistent estimator for the true value $\sbtheta$ of the parameter vector for the partial model \eqref{eq:parmod}. In particular, Proposition~\ref{prop:allentro} holds with $\sbtheta$ replaced by $\hbtheta(\bX^n,\bY^n)$.

We now define a \emph{nested null model}, with the object of testing the null hypothesis that the TE \eqref{eq:trent} is zero. Assuming the partial model \eqref{eq:parmod} with parameter vector $\btheta$, it is clear that $\te Y X = 0$ iff $f(x_t|\bx^k_t,\by^k_t;\btheta)$ does not depend on $\by^k_t$. Accordingly, we \emph{define} the null set $\Theta_0 \subseteq \Theta$ by
\begin{equation}
	\Theta_0 \equiv \{\btheta \in \Theta \,|\, f(x_t|\bx^k_t,\by^k_t;\btheta) \text{ does not depend on } \by^k_t\} \,. \label{eq:nullset}
\end{equation}
We assume that $\Theta_0 \ne \emptyset$. Given a realisation $(\bx^n,\by^n)$ of the joint process $(\bX,\bY)$, the \emph{likelihood ratio test statistic} for the null hypothesis $H_0: \btheta \in \Theta_0$ is
\begin{equation}
	\Lambda(\bx^n,\by^n) \equiv \frac{\likel(\hbtheta_0|\bx^n,\by^n)}{\likel(\hbtheta_{\phantom 0}|\bx^n,\by^n)}
\end{equation}
where $\hbtheta$ and $\hbtheta_0$ are ML estimators for $\btheta$ over the full parameter set $\Theta$ and the null subset $\Theta_0$ respectively. The following Theorem justifies defining the \emph{model TE estimator}
\begin{equation}
	\tee Y X{\bx^n,\by^n} \equiv -\frac 1{n-k} \log \Lambda(\bx^n,\by^n) \,. \label{eq:tee}
\end{equation}
\begin{temltheo} \label{theo:tee}
\ \\ \vspace{-5mm}
\begin{itemize}
\item[(a)] \label{it:tee} $\tee Y X{\bX^n,\bY^n} \aslim \te Y X$ as $n \to \infty$.
\item[(b)] If $\te Y X = 0$, then $2(n-k)\tee Y X{\bX^n,\bY^n}$ has an asymptotic $\chi^2(d)$ distribution, where the number of degrees of freedom $d$ is the difference between the number of parameters in the full and null models. If $\te Y X > 0$, the asymptotic distribution is non-central $\chi^2(d;\lambda)$ with non-centrality parameter $\lambda = 2(n-k) \te Y X$. \qed
\end{itemize}
\end{temltheo}
\noindent Theorem~\ref{theo:tee}a follows immediately from Propositions~\ref{prop:allentro},\ref{prop:mle} and \resub{definitions \eqref{eq:trent} and \eqref{eq:allem}}; the estimator \eqref{eq:tee} is thus consistent, although it will generally be biased. Theorem~\ref{theo:tee}b follows from the standard large-sample theory \cite{Wald43} (but note that convergence to the non-central $\chi^2$ will generally be slower than in the null case). It enables significance testing of the null hypothesis of zero TE, and the construction of confidence intervals for the estimator if the sample is sufficiently large.


We note the ergodicity condition \eqref{eq:ergodic} restricts the class of models for which our analysis applies, although the restriction may not be that stringent. For example, for a (possibly nonlinear) VAR process of the form $X_t = g(\bX^k_t) + \veps_t$ with iid residuals $\veps_t$, the joint process $(X_{t-\ell},\ldots,X_t)$ will be ergodic for any lag $\ell$ if (roughly) for any region $\mathcal R \subset S_X$, $\prob{\veps_t \in \mathcal R} = 0$ $\implies \mathcal R$ has measure zero \cite{Doob53}. (This will be the case, for instance, if the residuals are nondegenerate multivariate Gaussian.) A discrete-valued Markov processes is ergodic if every state is aperiodic and positive recurrent \cite{Doob53}.


For a linear VAR partial model of the form $X_t = \sum_{i = 1}^k A_i X_{t-i} + \sum_{i = 1}^k B_i Y_{t-i} + \veps_t$ with iid multivariate Gaussian residuals $\veps_t$ with covariance matrix $\Sigma$ and regression coefficient matrices $A_i,B_i$, the ergodicity condition is met provided the \emph{generalised variance} \cite{Barrett:2010} $|\Sigma|$ is $> 0$. The parameter vector is $\btheta = (A_i,B_i,\Sigma)$ and the null set $\Theta_0$ is given by $B_1 = \ldots = B_k = 0$. The ML is (up to a factor) $|\hSigma|^{-(n-k)/2}$ where $\hSigma$ is the ML estimator for $\Sigma$, which we note is asymptotically equivalent to the conventional OLS estimator \cite{Hamilton94}. The TE estimator \eqref{eq:tee} is then just half the Granger causality from $\bY$ to $\bX$ \cite{Geweke82}, and we recover the result of \cite{Barnett:2009}.


Given a third process $\bZ$ jointly distributed with $(\bX,\bY)$, the effect of $\bZ$ on the information flow $\bY \to \bX$ may be ``conditioned out'' by defining the \emph{conditional transfer entropy} \cite{Schreiber00,Geweke84}
\begin{equation}
	\cte Y X Z \equiv \centro{X_t}{\bX^k_t,\bZ^k_t} - \centro{X_t}{\bX^k_t,\bY^k_t,\bZ^k_t} \,. \label{eq:ctrent}
\end{equation}
It may be verified that Theorem~\ref{theo:tee} extends to the conditional case for partial Markov models of the form
\begin{equation}
	p(x_t | \bx^{t-1},\by^{t-1},\bz^{t-1}; \btheta) = f(x_t|\bx^k_t,\by^k_t,\bz^{t-1};\btheta) \,. \label{eq:cparmod}
\end{equation}


In the discrete case, the various densities $p(\cdots)$ are actual probabilities. We may calculate from \eqref{eq:alle1} [for notational compactness we suppress explicit dependence on a realisation $(\bx^n,\by^n)$] that
\begin{equation}
	\allem(\btheta) = \sum_{\xi_t, \bxi^k_t,\beeta^k_t} \hp(\xi_t,\bxi^k_t,\beeta^k_t) \log f(\xi_t|\bxi^k_t,\beeta^k_t;\btheta) \,, \label{eq:alle2}
\end{equation}
where
\begin{equation}
	\hp(\xi_t,\bxi^k_t,\beeta^k_t) \equiv \frac 1{n-k} \sum_{t = k+1}^n \prod_{s = t-k}^t \delta(\xi_s,x_s) \prod_{u = t-k}^{t-1} \delta(\eta_u,y_u) \label{eq:pmle}
\end{equation}
are plug-in estimates for the densities $p(x_t,\bx^k_t,\by^k_t)$ (by the ergodic assumption they are are consistent). Eq.~\eqref{eq:alle2} then furnishes a direct route to calculation of the maximum-likelihood estimator $\hbtheta$ and thence the TE estimator \eqref{eq:tee}.

In the absence of an explicit model, the conventional non-parametric estimator for TE in the discrete case is the plug-in estimator
\begin{equation}
	\tefna Y X \equiv -\sum_{\xi_t, \bxi^k_t} \hp(\xi_t,\bxi^k_t) \log \hp(\xi_t|\bxi^k_t)
	+ \sum_{\xi_t, \bxi^k_t,\beeta^k_t} \hp(\xi_t,\bxi^k_t,\beeta^k_t) \log \hp(\xi_t|\bxi^k_t,\beeta^k_t) \label{eq:trentf}
\end{equation}
(with the obvious notation). Now in the case of a \emph{finite} state space, we can consider the transition probabilities $p(\xi_t|\bxi^k_t,\beeta^k_t)$ \emph{themselves} as model parameters; \ie\ the $\btheta$ in a partial Markov model \eqref{eq:parmod}. Then it is clear that the $\hp(\xi_t|\bxi^k_t,\beeta^k_t)$ will be ML estimators, and the plug-in estimator \eqref{eq:trentf} is just the model estimator \eqref{eq:tee}. If the cardinalities of the state spaces $S_X,S_Y$ are $a,b$ respectively, then there are $(a-1)a^k b^k$ independent parameters for the non-null model and $(a-1)a^k$ for the null model. Theorem~\ref{theo:tee}b then states that the plug-in estimator has an asymptotic $\chi^2$ distribution with $(a-1)a^k(b^k-1)$ degrees of freedom. Note that this scales polynomially with state space size and exponentially with model order.

This is illustrated in the following example of a stationary, bivariate first-order Markov chain on the binary state space $\{0,1\} \times \{0,1\}$. \resub{Although manifestly a toy model, we present this example partly for its close resemblance to a canonical minimal model analysed in \cite{KaiserSchreiber02}, but mostly as it illustrates nicely the asymptotic convergence of the TE estimator to the theoretical $\chi^2$ distribution.} For $t = 1,2,\ldots$, $\veps_t$ and $\eta_t$ are iid $B(\shalf)$ random variables \footnote{$B(p)$ denotes a Bernoulli trial taking the value $1$ with probability $p$.} and $u_t,v_t$ are iid $B(\theta), B(\vphi)$ respectively. We then define
\begin{equation}
\begin{split} \label{eq:mm}
	X_t &= u_t Y_{t-1} + (1-u_t) \veps_t \\
	Y_t &= v_t X_{t-1} + (1-v_t) \eta_t \,.
\end{split}
\end{equation}
It is clear that the ergodic requirement is satisfied provided $\theta < 1$ and $\vphi < 1$. We may calculate
\begin{equation}
	p(x_t,y_t|x_{t-1},y_{t-1}) = p(x_t|x_{t-1},y_{t-1})  q(y_t|x_{t-1},y_{t-1}) \label{eq:mmf}
\end{equation}
where the conditional marginals are
\begin{equation}
\begin{split}
	p(x_t|x_{t-1},y_{t-1}) &= \theta \delta(x_t,y_{t-1}) + \shalf (1-\theta) \\
	q(y_t|x_{t-1},y_{t-1}) &= \vphi  \delta(y_t,x_{t-1}) + \shalf (1-\vphi) \,.
\end{split}
\end{equation}
 Note that in this case the extended model \eqref{eq:fullmod} is indeed misspecified, since from \eqref{eq:mm} it can be seen that $Y_t$ is \emph{not} independent of $X_t$ conditioned on joint history; the joint conditional pdf \eqref{eq:mmf} does not factor according to \eqref{eq:fullmod}.

\begin{figure}
\begin{center}
\includegraphics{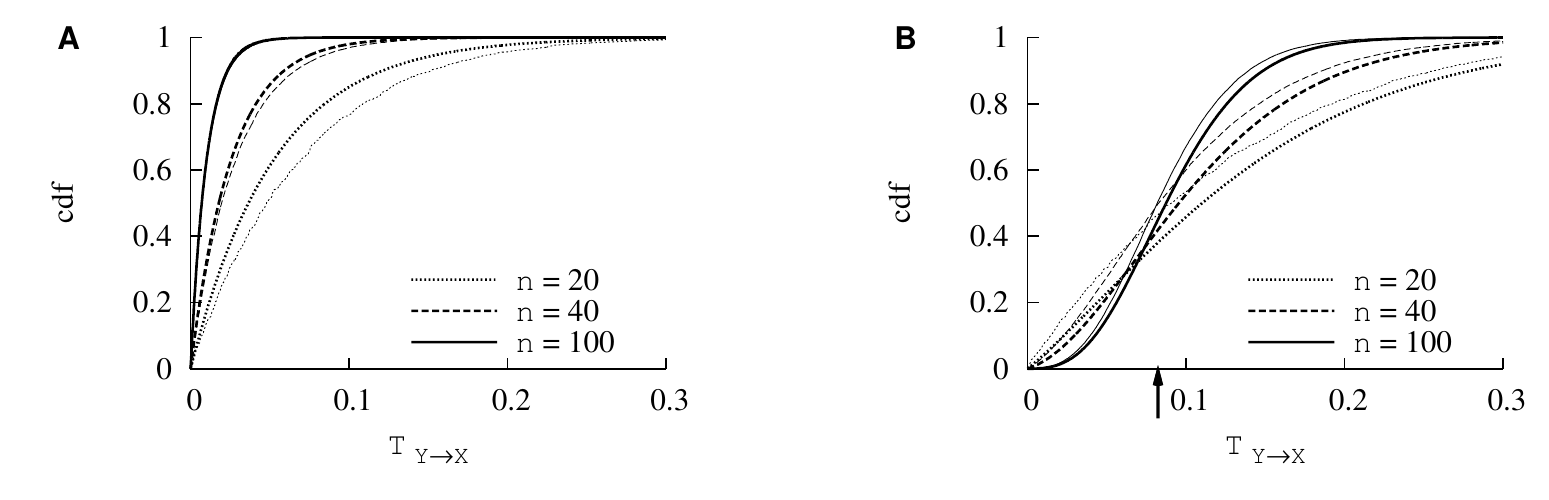}
\end{center}
\caption{Cumulative distribution functions (cdfs) for the plug-in TE estimator \eqref{eq:trentf} and corresponding scaled $\chi^2(2)$ theoretical asymptotic distributions (bold lines), for \small{\textbf{\textsf{A}}} a null model ($\theta = 0$) and \small{\textbf{\textsf{B}}} a non-null model ($\theta = 0.4$) for the binary Markov process \eqref{eq:mm} at a series of sequence lengths $n$. The vertical arrow marks the actual TE value $\te Y X \approx 0.0823$.} \label{fig:mm}
\end{figure}

By the $x \leftrightarrow y$,$\theta \leftrightarrow \vphi$ symmetry, it follows that the stationary joint density is uniform $p(x,y) = \frac 1 4$ for all $x,y$, and we may calculate
\begin{equation}
	\te Y X = \shalf (1+\theta)\log(1+\theta)+\shalf(1-\theta)\log(1-\theta) \,. \label{eq:mmte}
\end{equation}
By Theorem~\ref{theo:tee} the TE plug-in estimator \eqref{eq:trentf} in this case has an asymptotic $\chi^2(2)$ distribution under the null hypothesis of zero TE, and a non-central $\chi^2(2)$ distribution under the alternative hypothesis. \figref{fig:mm} plots empirical vs. theoretical $\chi^2(2)$ cumulative distributions (cdfs) of $\tefna Y X$ for a null model ($\theta = 0$, \figref{fig:mm}A) and a non-null model ($\theta = 0.4$, \figref{fig:mm}B), estimated from $10^5$ realisations each of a selection of sequence lengths. The $\vphi$ parameter was set to $0.6$.
We see that the null distribution converges more quickly to its $\chi^2(2)$ asymptote than the non-null distribution.

\resub{With regard to application of the finite-state plug-in estimator result, we note that it is not uncommon to \emph{discretise} continuous time series data for causal analysis, sometimes after a differencing step. A more sophisticated discretisation approach is the recently introduced \emph{symbolic transfer entropy} \cite{StaniekLehnertz:2008}, currently gaining in popularity as a technique for inference of time-directed information flow applicable to systems of continuous variables.}


\resub{A recognised problem with transfer entropy estimation in the non-parametric case is the so-called ``curse of dimensionality'' \cite{RungeEtal:2012}, in particular with regard to scaling of degrees of freedom with number of lags. Thus for continuous processes, naive coarse-graining and binning of data points is likely to be extremely inefficient in practice, yielding large estimation errors; worse, estimators derived in this fashion may not even converge monotonically to the correct value \cite{KaiserSchreiber02}. There are several approaches to mitigation of this problem, including adaptive partitioning, kernel density methods \cite{KaiserSchreiber02} and $k$-nearest neighbour statistics \cite{KraskovEtal:2004}. An interesting avenue of research is whether any of these methods might be framed in a parametric context for which our result applies, thus yielding a useful $\chi^2$ sampling distribution. Another promising approach is proposed in \cite{RungeEtal:2012}, where transfer entropy is decomposed into contributions of individual lags and then conditional dependencies reduced using methods from graphical model theory \cite{Lauritzen:1996}. Again, it would be of interest to investigate whether a similar dimensional reduction could be achieved within a model-based ML framework.}

\resub{In a parametric context, it might be argued that dimensionality is a lesser curse: prior to parametric TE estimation an empirical model order should be selected in accordance with the size of the available sample, most conveniently, as noted previously, via the average log-likelihood estimator \eqref{eq:alle1}. In practice this will limit the dimensionality of the parameter space and will frequently yield tractable model orders, resulting in acceptably efficient estimators. Thus for example, Granger causality has been effectively applied (especially in the neurosciences) for highly multivariate data. For discrete systems, and in particularly for symbolic transfer entropy, it seems not unreasonable to expect similar, although further research is required.}

\resub{The special case of Granger causality suggests a useful range of applications for Theorem~\ref{theo:tee}. Linear VAR modelling seems sometimes to be taken as a convenient ``one size fits all'' approach, insofar as (almost) any wide-sense stationary process may be modelled as a linear VAR, albeit of potentially high model order \cite{Hamilton94,Geweke82}. However, this is certainly not to say that a linear VAR model will necessarily be a \emph{good} model for given time series data. In particular, high empirical model orders (as indicated by model selection criteria) may well be indicative of failure of a model to reflect parsimoniously the statistical structure of the data. Physical considerations or supplementary analysis may suggest alternative models. Thus for example a \emph{nonlinear} VAR approach \cite{Ancona:2004,ChenEtal04} might be suspected on physical grounds, heteroscedasticity may suggest a GARCH model \cite{LuoEtal:2011} or long-term memory a fractional ARIMA model \cite{GrangerJoyeux:1980,Hoskin:1981}. Now it may be far from clear how one should define a Granger-like predictive statistic for such models. Theorem~\ref{theo:tee}, and its implied equivalence (at least in the Gaussian case) with transer entropy, suggests that the principled generalisation of Granger causality is just the model transfer entropy, operationalised as a log-likelihood ratio - and in addition furnishes an asymptotic sampling distribution.}

In summary, our result unifies the established machinery of maximum likelihood and the concomitant large-sample theory with the information-theoretic notion of directed information transfer. Given a---perhaps physically motivated---parametric model (or, if the state space is discrete, an implicit Markov chain model) it thereby facilitates estimation and statistical inference of transfer entropy. In particular it furnishes generic asymptotic $\chi^2$ distributions for significance testing and estimation of confidence intervals, obviating the need for surrogate/subsampling methods.

Finally, current research by the authors suggests an extension of the result to \emph{point processes}, with application to inference of information flow in spiking neural systems \cite{OkatanEtal:2005,KimEtal:2011}.

Author Barnett's work has been supported by the Dr. Mortimer and Theresa Sackler Foundation via the Sackler Centre for Consciousness Science, and a visiting fellowship at the Centre for Research in Complex Systems.

\bibliography{teml}

\end{document}